\documentclass[prf,
reprint,
amsmath,amssymb,twocolumn
]{revtex4-2}
\usepackage{graphicx}
\usepackage[english]{babel}
\usepackage{amssymb}
\usepackage{amsbsy}
\usepackage{amsmath}
\usepackage{hyperref}

\newcommand{\cond}{K}

\newcommand{\R}{\text{Re}}

\begin{document}
\title{A dual network approach to connect structure and flow in random networks}
 \author{Marco Dentz$^1$, Philippe Gouze$^2$, Tanguy le Borgne$^3$, Alexandre Puyguiraud$^1$}
 \affiliation{$^1$Spanish National Research Council (IDAEA-CSIC),
 Barcelona, Spain}
 \affiliation{$^2$Geoscience Montpellier, CNRS, Universit\'e de
 Montpellier, Montpellier, France}
 \affiliation{$^3$Geoscience Rennes, Universit\'e de Rennes, Rennes,
 France}

\begin{abstract}
Flows in random networks are known to exhibit heavy-tailed statistics,
giving rise to anomalous transport in a range of biological,
environmental and engineered systems. Yet, how structure determines flow
remains an open question. Here we derive a dual network
approach that relates flow statistics and network
properties. Conditional statistics in a range of networks reveal the
existence of two interlaced subnetworks with distinct hydraulic
behaviors. Based on this hidden structure, we derive a universal
analytical approach that predicts heavy-tailed flow statistics based on network
topology and disorder distribution across a range of random
networks.

\end{abstract}
 \maketitle


Flows in random networks are ubiquitous over a range of
scales and disciplines,  including fluid flow in pore and vascular
networks, fracture, karst and river networks, as well as current
distributions in resistor networks \cite{BG1990,Sahimi}. Naturally
occurring and engineered networks are inherently heterogeneous,
characterized by broad distributions of
spatially fluctuating conduit diameters and conductances.
While flow distributions in random networks have been systematically
observed to follow heavy-tailed statistics, the relation between
network structure and flow statistics remains an open question. This
knowledge gap is significant because the dispersion of solutes,
nutrients and active and passive matter are determined by the flow
distribution rather than the mean flow. Flow heterogeneity
leads to anomalous transport behaviors that manifest in broad residence and
transit time distributions, and preferential transport
pathways
\cite{Bruderer2004relating,Bijeljic2011,hyman2019emergence,goirand2021network,davy2024structural}.
Current anomalous transport approaches generally need to assume or
measure flow statistics rather than deriving them directly from the
medium structure \cite{LBDC2008:2,Dentz2016,hyman2019emergence,goirand2021network}.
Furthermore, extreme flow behaviors, emerging flow patterns,
bypass and filtration behaviors, dominant flow paths and bottlenecks cannot be
captured by measures for the mean flow. However, it is these features
that dominate network scale flow and transport behaviors and the occurence of
extreme events.

\begin{figure}
\includegraphics[width=0.45\textwidth]{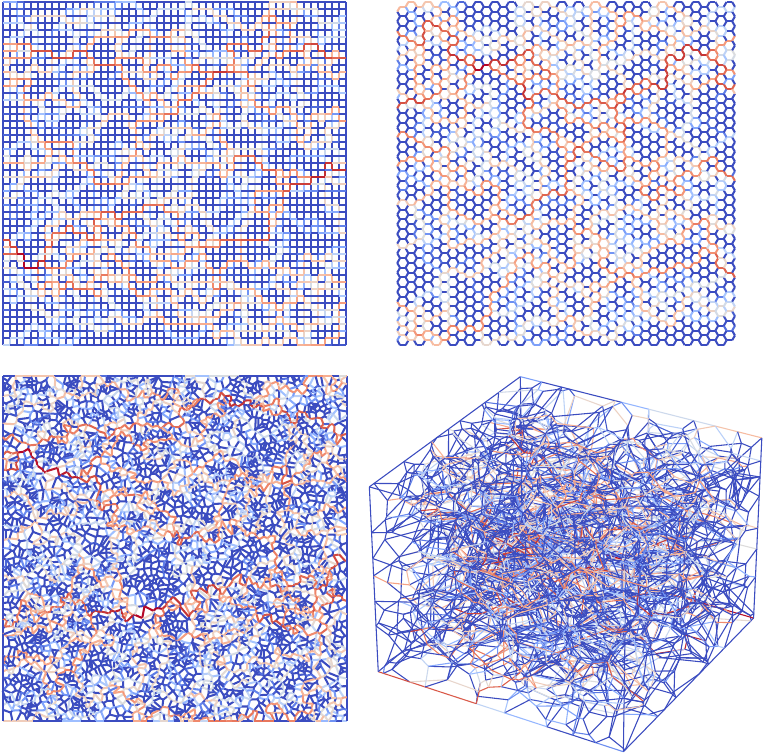}
\caption{Illustration of flow rate distributions in two- and
three-dimensional random networks for $\alpha = 1/2$. The top row
shows the regular (left) square and (right) honeycomb networks with $d
= 3$ and $4$ and identical diameter distribution. The bottom panel
shows the (left) 2D and (right) 3D random Voronoi networks with $d
= 3$ and $4$ and identical diameter distribution. Colors are
logarithmically scaled from (blue) small to (red) large values.
The dark blue conduits denote flow rates smaller than $10^{-3}$ times
the mean flow rate. \label{fig:maps}}
\end{figure}

Flow velocities in pore and fracture networks are typically broadly
distributed with behaviors that can be characterized by lognormal and
Gamma-type distributions with power-law scalings at low velocities and
exponential or stretched exponential decay at velocities much larger
than the mean
\cite{Holzner2015,aramideh2018pore,souzy2020velocity,davy2024structural}. Similar
behaviors have been found for current distributions in random resistor
networks at the percolation threshold and for broad conductance distributions
\cite{duering1992current, sena2023decomposing}. In the context of
porous networks, models to quantify these
behaviors assume a constant pressure drop across conduits, which
allows mapping the diameter distribution onto the distribution of flow
rates via the Poiseuille law
\cite{saffman1959,deAnna2017,puyguiraud2021pore}. The work by
\citet{alim2017local}, on the other hand, emphasizes the role of network nodes in the focusing and
redistribution of flow, and adapts the $q$-model for force
distributions in granular media \cite{coppersmith1996} to explain exponential flow rate
distributions in weakly heterogeneous porous media. Percolation theory
and critical path analysis have been used to estimate the effective
conductivity, that is, the mean flow,
in random networks \cite{hunt2017flow}. These approaches
are able to describe certain aspects of network flow, but fall short
in bridging the gap between network structure and flow
distribution. 

In this letter, we derive a theory that establishes the relation
between network structure and flow statistics. Conditional flow
statistics across different network structures and heterogeneity
reveal the existence of two types of interlaced flow networks with
distinct hydraulic behaviors. This hidden dual structure forms the basis for the derivation of an
analytical model to explain how diameter distribution and connectivity
determine the distributions of flow rates and flow velocities.


\begin{figure}[t]
\centering
\includegraphics[width=0.48\textwidth]{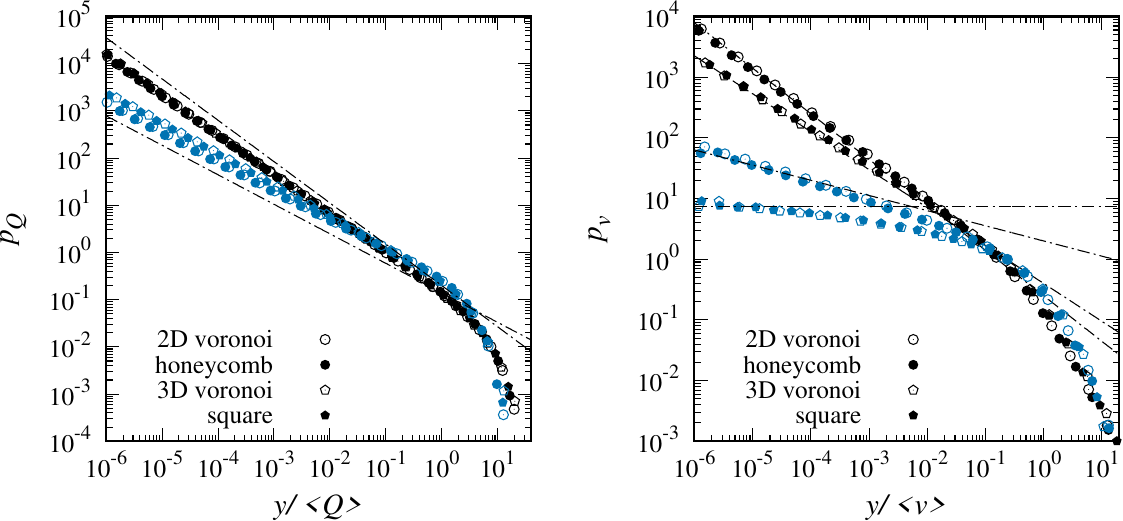}
\caption{Distributions of (left) flow rate and (right) velocity for
(top panel) regular and (bottom panel) irregular networks for (black)
$\alpha = 1/2$ and (blue) $\alpha = 3/2$. The dash-dotted
denote (right) the scalings $p_Q(y) \propto y^{\beta
-1}$ with $\beta = 0.125$ and $0.375$, (left) the scalings $p_v(y)
\propto y^{\gamma - 1}$ with (top
to bottom) $\gamma = 0.25, 0.375, 0.75, 1$. \label{fig:Q}}
\end{figure}


The network structure is
characterized by connectivity (node degree) and the distribution of conduit
diameters, or equivalently, conductances. The networks consist of nodes and
conduits that connect adjacent nodes. We consider two regular
network types, honeycomb with node degree $d = 3$, and square  ($d =
4$), and two irregular networks, two-dimensional
(2D) Voronoi networks ($d = 3$), 3D Voronoi networks ($d = 4$), see
Figure \ref{fig:maps}. These
types of networks have been used as representations
of vascular and porous networks \cite{karst2017model} and abstractions
for fractured and karstic networks \cite{hyman2017predictions,
collon2017statistical}. Additional numerical simulations for a Berea
sandstone pore network are reported in the Appendix. 

Network flow is linear and determined by the Poiseuille law,
which relates pressure drop $\Delta P_{ij} = P_{i} - P_j$ and flow rate
$Q_{ij}$ in the conduit connecting nodes $i$ and $j$ as $Q_{ij} =
\kappa_{ij} \Delta P_{ij}$. The conductance is $\kappa_{ij} = \pi D^4/128
\mu \ell_0$, where $\mu$ is dynamic viscosity. The flow velocity is $v_{ij}
= {Q_{ij}}/{A_{ij}}$ with $A_{ij} = \pi D_{ij}^2/4$ the cross-sectional
area of the conduit. The conservation of fluid mass
is described by the continuity equation $\sum_{[ij]} Q_{ij} = 0$,
where $\sum_{[ij]}$ denotes the sum over all nodes $j$ connected to
node $i$. The Kirchhoff problem is solved with constant pressure
boundary conditions at the left and right boundaries, corresponding to
a unit pressure gradient. The conduit diameters $D$ are distributed
according to the truncated Gamma distribution
$p_D(x) \propto \left({x}/{D_0} \right)^{\alpha -1} {\exp(-x/D_0)} 
$
for $D/D_0 > 10^{-5}$. Thus, the distribution $p_\kappa(k)$ of
conductance behaves as $p_\kappa(k) \propto \left({k}/{\kappa_0}\right)^{\alpha/4 - 1}
\exp[-({k}/{\kappa_0})^{1/4}]$,
where
$\kappa_0 = {\pi D_{0}^4}/{128 \mu \ell_0}$. The full numerical setup
is given in the Appendix. 

\begin{figure*}
\includegraphics[height=0.28\textwidth]{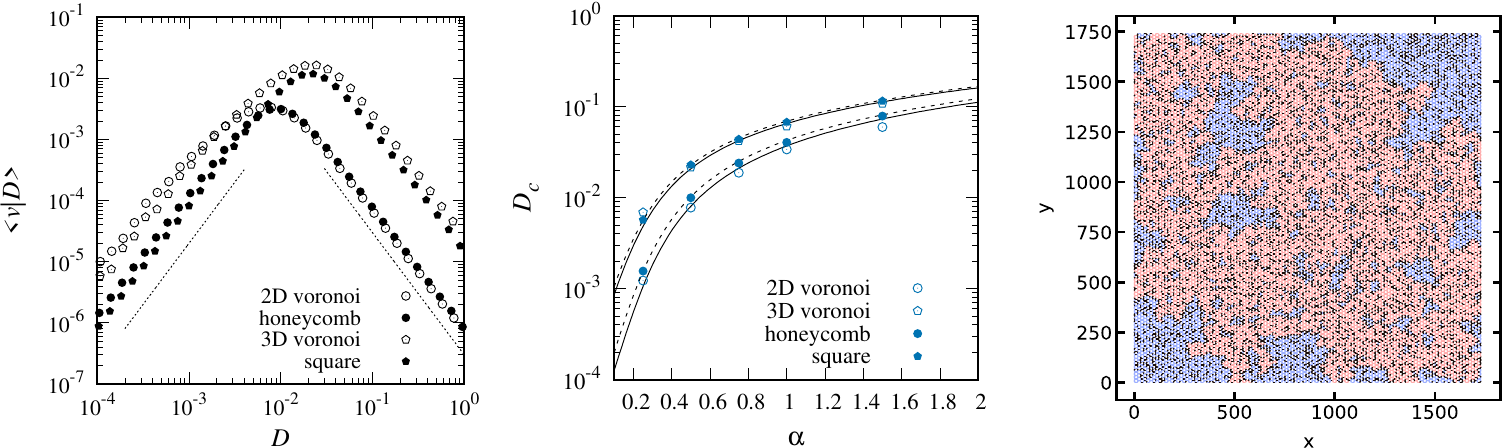}
\caption{Left: Conditional average of flow speed for the (full symbols) regular and (open symbols) irregular
networks as functions of the conduit diameter for $\alpha =
1/2$. The dashed lines indicate the scalings $\langle v|D \rangle \propto
D^{\pm 2}$. Center: Dependence of the critical diameter $D_c$ on the exponent
$\alpha$, obtained from the (symbols) maximum of $\langle v|D
\rangle$ and (lines) from the analytical relationship
\eqref{eq:pcdc}, see also the Appendix. 
Right: (Red) Spanning cluster
\cite{stauffer2018introduction}, (blue) isolated clusters, (black)
conduits with $D < D_c$ for a honeycomb network. Dangling ends drain
the conduits with $D < D_c$.\label{fig:cond}}
\end{figure*}

Many classical and modern studies on structure-flow relation focus on the analysis of
the effective conductance \cite{BG1990, hunt2017flow}, which characterizes the average flow
of the network. For the understanding of extreme flow behaviors and
anomalous dispersion, however, we require the full distributions of
flow rate and velocity. The distribution of flow rates is obtained by
sampling $Q_{ij}$ over all conduits, 
\begin{align}
p_Q(y) = \frac{1}{N_c} \sum_i \sum_{[ij)} \delta(y - |Q_{ij}|),
\end{align}
where $\sum_i$ denotes the sum over all nodes, $\sum_{[ij)}$ the sum over all
downstream connections from node $i$, $N_c$ is the number of all
conduits, $\delta(\cdot)$ denotes the Dirac-delta. Network transport is determined by the distribution of
the flow velocities $v_{ij}$ between nodes. The Lagrangian velocity
distribution, that is, the distribution of particle velocities along
pathlines can be obtained by volume-weighted sampling of $v_{ij}$ over
all conduits (see Appendix) 
\begin{align}
\label{eq:pv}
p_v(y) = \frac{1}{V}\sum_i \sum_{[ij)} V_{ij} \delta(y - |v_{ij}|)
\end{align}
where $V_{ij} = \pi D_{ij}^2 \ell_{ij}/4$ is the conduit volume and
$V = \sum_i \sum_{[ij)} V_{ij}$ is the total volume. Understanding $p_v(y)$ provides the
link between structure and transport. Figure \ref{fig:Q} shows the
distributions of flow rate and flow velocity for the six
network topologies under consideration with $\alpha = 1/2$ and
$3/2$. The flow rate and flow speed distributions behave as power laws
$p_Q(y) \propto y^{\beta -1}$ and $p_v(y) \propto y^{\gamma - 1}$ for
values much smaller than the mean. The results show that the flow rate
distributions are independent from the network topology and
depend only on the diameter distribution. The speed distribution
$p_v(y)$, in contrast, depends strongly on the network type, and,
specifically, on the node degree: The larger the node degree, the
larger larger the exponent $\gamma$, with an apparent saturation at
$\gamma = 1$ for $d = 4$ and $\alpha = 1.75$.

In order to understand and quantify these behaviors in the light of the network
structure, we focus now on the mean flow speed $\langle v|D \rangle$
conditioned on the conduit diameter. Figure \ref{fig:cond}
shows the conditional averages for the four network types under
consideration and $\alpha = 1/2$. The observed behaviors are analogous
for different values of $\alpha$ and in fact for other diameter
distributions. The conditional mean speed scales as $\langle v|D
\rangle \propto D^{2}$ for $D$ smaller than a critical diameter $D_c$
and $\propto D^{-2}$ for $D \geq D_c$. The behaviors for the
conditional mean flow rate and pressure drop are given in the
Appendix. 

The critical diameter $D_c$ depends on the network type and increases
as the node degree increases, see Figure \ref{fig:cond}. It separates
two subnetworks of distinct flow behaviors. The subnetwork with $D
\geq D_c$ behaves in average like a system in series, and its
complement as a system in parallel \footnote{In a serial arrangement
of conduits, the flow rate is
constant due to mass conservation, and the pressure drop varies
according to the conduit diameters, while
for a system of parallel conduits, the pressure drop is constant and
the flow rate varies. This behavior is true here only in
average.}. However, this analogy is only true in average. The
mechanisms that give rise to these behaviors are different from the
ones for serial or parallel configurations. 

First, we delineate the subnetworks by considering the fraction
$p_c$ of the network conduits that are populated by diameters larger
than the critical diameter $D_c$
\begin{align}
\label{eq:pcdc}
p_c = \int\limits_{D_c}^\infty d x p_D(x).
\end{align}
We find that the value of $p_c$ is independent of the diameter distribution, and is
a characteristic of the underlying network. Figure \ref{fig:cond} shows
$D_c$ as a function of the exponent $\alpha$ obtained from the maximum
of $\langle v|D \rangle$ \cite{SI} and from relation \eqref{eq:pcdc} for
constant $p_c = 0.653$ (honeycomb), $0.5$ (square), $0.69$ (2D
Voronoi), $0.52$ (3D Voronoi). These
values are consistent with the bond percolation thresholds for the
respective networks (honeycomb, square, and 2D Voronoi)
\cite{stauffer2018introduction,becker2009percolation}, see also the Appendix. This
remarkable finding indicates that the conduits with $D \geq D_c$ form
the percolation clusters of the respective networks, see Figure
\ref{fig:cond} for the spanning cluster of a honeycomb network.    
It is interesting to note that critical path analysis and continuum
percolation use expressions similar to \eqref{eq:pcdc} to estimate the
network conductivity \cite{hunt2017flow}. Our results reveal that
the percolation threshold divides the network into two
hydraulically different subnetworks. 

\begin{figure*}
\centering
\includegraphics[height=0.28\textwidth]{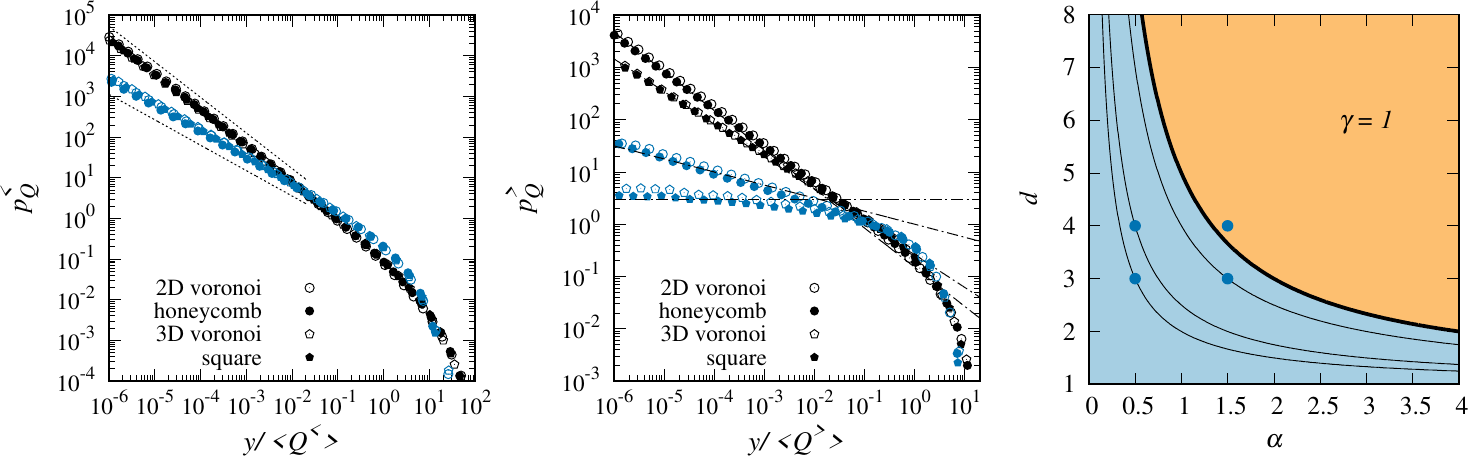}
\caption{Flow Rate and speed distributions for (black) $\alpha =
1/2$ and (blue) $\alpha = 3/2$. Left: Flow Rate distributions sampled in the conduits with $D
< D_c$. The dotted lines indicate the scaling $y^{\alpha/4-1}$. Center:
Flow Rate distributions for $D \geq D_c$. The solid lines indicate the
scalings $y^{(d-1)\alpha/4-1}$ for $(d-1)\alpha/4 \leq 1$ and $y^0$
for $(d-1)\alpha/4 > 1$. Right: Diagram of
the dependence of the exponent $\gamma = d \alpha/4$ of the speed distribution  
as a function of node degree $d$ and exponent $\alpha$ of the
diameter distribution. The thin black lines denote (left to right)
$\gamma = 0.25, 0.375, 0.75$, the symbols denote the
values corresponding to the center panel.\label{fig:Qcond}}
\end{figure*}
Accordingly, the global flow rate distribution can be decomposed into
\begin{align}
p_Q(y) = (1 - p_c) p^<_{Q}(y) + p_c p^>_{Q}(y). 
\end{align}
The superscripts $\gtrless$ denote quantities on the $D \geq D_c$ and
$D < D_c$ subnetworks.
The pressure drop is indeed variable between conduits
for $D < D_c$ unlike for a system of conduits in parallel. However, its
conditional distribution is the same for all conduit diameters, and
equally for the flow rate distribution at $D > D_c$, 
\begin{align}
p^<_{\Delta P|D}(z) = p^<_{\Delta P}(z), && p^>_{Q|D}(y) = p^>_Q(y),
\end{align}
%
see Appendix and Ref. \cite{kolek2000current}. 
To understand the behavior for $D < D_c$, we consider the pressure
drop across a conduit as given by the Dyson equation \cite{Sahimi2003,
bhattacharjee2023green}
\begin{align}
\label{eq:dyson}
\Delta P_{ij} = \Delta \Pi_{ij} + \sum\limits_{k,\ell}
\kappa'_{k\ell} (G_{ik} - G_{jk})\Delta P_{k\ell},
\end{align}
where $\kappa_{k\ell}'$ are the conductance fluctuations, $G_{ik}$
is the lattice Green function, and $\Delta \Pi_{ij}$ is the presssure
drop across the corresponding homogeneous network. For weak disorder,
\eqref{eq:dyson} can be linearized, which gives a Gaussian
distribution for $p^<_{\Delta P}(z)$, see Appendix. 
In general, the non-local
dependence of the pressure drop on the conductance fluctuations
indicates that $\Delta P_{ij}$ is only weakly dependent on the local
conductance consistent with our observations. The non-locality
increases with the disorder strength, that is, for decreasing
$\alpha$. Thus, the scaling of
$p^<_Q(y)$ for $y \ll \langle Q \rangle$ can be obtained directly from
the diameter or equivalently the conductance distribution as (see Appendix) 
\begin{align}
p^<_Q(y)
\propto y^{\alpha/4 - 1} \int\limits_{0}^\infty dz p_{\Delta P}(z)
z^{-\alpha/4},
\label{eq:scaling1}
\end{align}
for $\alpha/4 \leq 1$ and $p^<_Q(y) \propto y^0$ for $\alpha/4 > 1$
\footnote{The integral on the right side of
\eqref{eq:scaling1} exists because $p_{\Delta p}(z)$ decays slower than
$z^{\alpha/4-1}$, see also the Supplementary Material}, as shown in 
Figure \ref{fig:Qcond}. It is independent of the network
connectivity, and determined by the diameter, or conductance
distribution only.  

The distribution $p^>_Q(y)$, in contrast, depends both on the node degree
and the conductance distribution as shown in
Figure \ref{fig:Qcond}. Its behavior at small $y \ll \langle Q \rangle$ can be
attributed to dangling ends in the $D \geq D_c$ clusters, which would
carry zero flux if the small conduits were closed. Here, however,
these dangling ends are connected to $d -1$ conduits with diameters $D <
D_c$. They are drained into the $D > D_c$ network with the
flow rate, 
\begin{align}
Q_0^> = \sum_{i = 1}^{d -1} Q_i^<, 
\end{align}
which determines the local flow rate distribution in the vicinity of the
dangling ends. The fact that $p^>_{Q|D}(y)$ is independent of $D$ can be
rationalized by constructing $p_Q^>(y)$ as 
\begin{align}
\label{eq:pQl}
p_Q^>(y) = \int\limits_0^\infty dz p_0^>(z) z^{-1} f_Q(y/z), 
\end{align}
where $f_Q(z)$ denotes the flow rate distribution in response to a
unit injection and $p^>_0(y)$ is the distribution of $Q_0$. The latter
is obtained by the $(d-1)$--fold
convolution of $p^<_Q(z)$, which implies that $p_0^>(y) \propto
y^{(d-1) \alpha/4-1}$ for $y \ll \langle Q \rangle$. Thus, we obtain
the scalings (see Appendix) 
\begin{align}
\label{eq:pQ_large}
p_Q^>(y) \propto y^{(d-1) \alpha/4-1},
\end{align}
for $(d-1)\alpha/4 \leq 1$ and $p_Q^>(y) \propto y^0$ for $(d -
1)\alpha/4 > 1$, which are the behaviors shown in Figure
\ref{fig:Qcond}. The distribution $p^<_Q(y)$ dominates over $p^>_Q(y)$
at $y \ll \langle Q \rangle$. Thus, the exponent $\beta$ of $p_Q(y)$
is $\beta = \alpha/4$ for $\alpha \leq 4$ and $\beta = 1$ for $\alpha
> 4$. It scales in the same way as the conductance distribution. 

With these insights, we can now quantify the scaling of the velocity
distribution $p_v(y)$. First, we note that $p_v(y)$
is dominated by the conduits $D \geq D_c$ due to the volume weighting in
\eqref{eq:pv}. Second, the flow velocity is related to the flow rate by $v_{ij} = Q_{ij}/\pi
D_{ij}^2 \equiv Q_{ij}/A_{ij}$ while the flow rate distribution for $D
> D_c$ is independent of $D$. Therefore, $p_v(y)$ can be written as
\begin{align}
p_v(y) 
= \frac{1}{A_0}\int\limits_0^\infty da p^>_Q(y a) a^2 p_A(a),
\end{align}
where $A_0$ is the total cross-sectional area for $D \geq D_c$ and
$p_A(a)$ is the distribution of $A_{ij}$. This
implies that the power-law $p^>_Q(y) \propto y^{(d-1)\alpha/4 - 1}$ observed for
the flow rates at $D \geq D_c$ transfers to the velocity PDF such that
$p_v(y) \propto y^{(d-1)\alpha/4 - 1}$. That is, the exponent $\gamma$
is $\gamma = (d-1)\alpha/4$ for $\alpha\leq 4/(d-1)$ and $\gamma = 1$
for $\alpha\leq 4/(d-1) > 1$. These theoretical scalings capture
heavy-tailed statistics for small flow speeds observed across the
regular and random networks shown in Figure \ref{fig:Q} based only on
the node degree and diameter distribution. Figure \ref{fig:Qcond}
illustrates the dependence of $\gamma$ on network topology and
heterogeneity through the node degree $d$ and the diameter
distribution and delineates the different scaling regimes. An application of the
theory to a Berea sandstone pore network is given in the Appendix. 

In conclusion, our study provides a theoretical approach to connect
network poperties and flow statistics. Conditional statistics reveal
the existence of a dual network structure of different hydrodynamic
behaviors that are delineated by the networks' bond percolation
thresholds. These insights deliver a predictive theory for the flow
statistics in terms of the network properties. Our results have direct
implications for the understanding and prediction of flow, transport
and reaction patterns in natural and engineered heterogeneous media
from the micro to the kilometer
scale, including, porous, fractured media, as well as karst, river and
vascular networks. They can provide the basis for the design of
materials with predefined flow, and thus transport and mixing
properties. Due to the dominant role of structure and
heterogeneity, we expect similar behaviors for
non-linear network flows, such as turbulent and non-Newtonian.   

\noindent
{\bf Acknowledgments}

M.D. acknowledges funding by the European Union (ERC, KARST, 101071836).
\bibliography{ctrw}

\appendix

\onecolumngrid

\section{Numerical simulations}

The graph Laplacian or Kirchhoff matrix is a representation of the graph
underlying the network. The Laplacian here is weighted by the conductances of the
bonds of the graph. The conductance or weight matrix $\mathbf \cond$ contains the
conductances $\cond_{ij}$ between sites $i$ and $j$. If all non-zero
conductances are unity, $\mathbf \cond$ is equal to the adjacency matrix of the
graph. The weighted degree matrix is a diagonal matrix that contains the sum of
the conductances of the bonds adjacent to a node, that is, $D_{ij} = \delta_{ij}
\sum_{[ki]} \cond_{ik}$, where $\sum_{[ki]}$ denotes summation over all sites $k$
that are connected to site $i$. With these definitions, the weighted graph
Laplacian is given by $\mathbf L = \mathbf D - \mathbf K$. The continuity
equation for single-phase flow through the network can then be written as
\begin{align}
\label{eq:K1}
\mathbf L \cdot \mathbf P = \mathbf 0,
\end{align}
where $\mathbf P$ is the vector of pressures at all sites of
the network. This can be seen by defining the flow rate from site $i$ and $j$ as 
\begin{align}
    Q_{ij} = \cond_{ij} (P_i - P_j),  
\end{align}
which is a statement of momentum conservation. Note that $Q_{ij} = - Q_{ji}$.  
Continuity is expressed by
\begin{align}
\label{eq:mass}
    \sum_{[ji]} Q_{ij} = 0.
\end{align}
The latter is equivalent to Eq.~\eqref{eq:K1}. 
From the mass conservation statement~\eqref{eq:mass}, it follows that the graph
Laplacian has the eigenvector $(1,\dots,1)$ with the eigenvalue zero, which is
the trivial solution of equal pressure at all sites. Thus, the determinant of
the graph Laplacian is zero, i.e., $\mathbf L$ is not invertible. 

We seek a solution to Eq.~\eqref{eq:K1} for prescribed pressure at the
inlet and outlet nodes. To this end, we define the modified graph Laplacian
$\mathbf L'$ such that $L'_{ij} = \delta_{ij}$ for all sites $i,j$ that are in
the inlet or outlet boundaries and $L'_{ij} = L_{ij}$ for all other sites. With
this definition, the boundary value problem can be written as
\begin{equation} \label{eq:K2}
{\mathbf L}^\prime \cdot \mathbf P = \mathbf b,
\end{equation}
where the vector $\mathbf b$ on the right side of Eq.~\eqref{eq:K2}
contains the boundary conditions with $b_{i} = p_0$ for the sites $j$ at the
inlet boundary and $b_i = 0$ else.

For laminar flow conditions, that is, $\R \ll 1000$, the conductance $\cond_{ij}$ are independent of the
flow rate and given only in terms of the hydraulic diameter and length of the
conduit according to the Poiseuille law,
\begin{align}
Q_{ij} = \frac{\pi D_{ij}^4}{128 \mu \ell_{ij}} (P_i - P_j),
\end{align}
where $D_{ij}$ is the hydraulic diameter of the conduit, $\ell_{ij}$ its length. Thus, the conductances are given by
\begin{align}
\label{eq:Gij_lin}
\cond_{ij} = \frac{\pi D_{ij}^4}{128 \mu \ell_{ij}}. 
\end{align}
The sparse linear system in Eq.~\eqref{eq:K2} is solved by direct
inversion. The numerical solution and data analysis is
implemented in Python, and uses the Python packages 
SciPy \cite{scipy} and NumPy \cite{numpy} in addition to NetworkX \cite{networkx} and OpenPNM
\cite{openpnm} for network generation, manipulation and analysis. The
networks used have on the order of $10^6$ nodes for all networks. Furthermore, statistics are
sampled over $10^2$ independent realizations of the diameter
distribution for the data shown in the main paper.  

\section{Berea network}

\begin{figure*}[t!]
\includegraphics[width=0.6\textwidth]{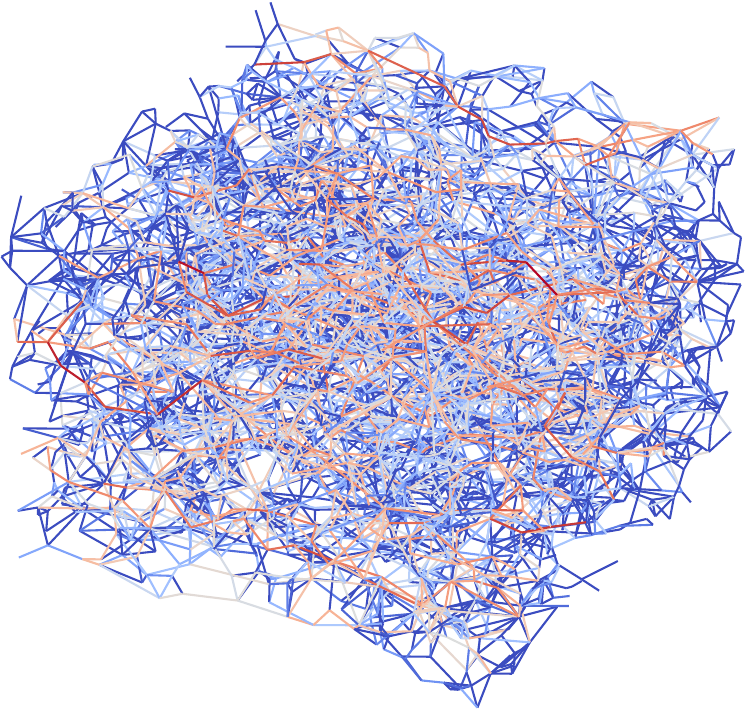}
\caption{Flow rate distribution in the pore network for a Berea sandstone from \citet{dong2009pore}.  
\label{fig_app:berea_map}}
\end{figure*}

\begin{figure*}[t!]
\includegraphics[width=0.4\textwidth]{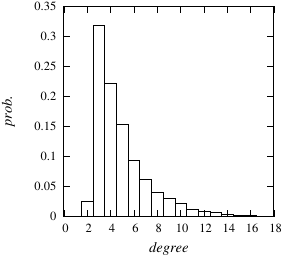}
\includegraphics[width=0.4\textwidth]{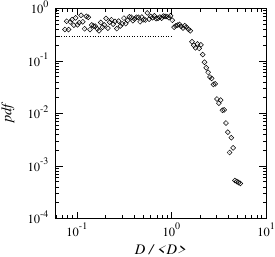}

\includegraphics[width=0.4\textwidth]{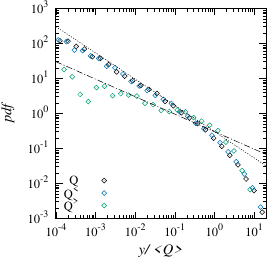}
\includegraphics[width=0.4\textwidth]{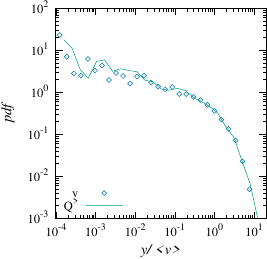}
\caption{The top panel shows the (left) distribution of node degrees
and (right) conduit diameters for the Berea network. The dominant node
degree is $d = 3$ with a small portion of nodes with degree $d = 2$
which are located at the boundaries. The bottom panel shows the 
distributions of (left) flow rates and (right) velocities for
the Berea network. The distributions follow the behaviors expected
from the dominant node degree $d = 3$ and diameter distribution. 
\label{fig_app:berea}}
\end{figure*}

The Berea network of \citet{dong2009pore} has been retrieved from the
website \citet{berea-data}. The conduits are connecting the pore
centers. The conductances of the conduits are obtained from the
diameters of the pore throats. Following \cite{blunt2017multiphase},
the conductances of the conduits are obtained from the throat
diameters only, because they dominate the hydraulic resistance compared
to the pore diameters. Figure \ref{fig_app:berea_map} shows the berea
pore network, and Figure \ref{fig_app:berea} shows the degree
distribution for the Berea network as well as the diameter
distribution, which is approximately exponential. 

The distributions of $Q^>$ and flow
velocities $v$ are dominated by the lowest node degree, which here
is $d = 3$ with a strong peak, see also Figure
\ref{fig_app:berea}. Nodes with degree $2$ are exclusively at the
boundaries. Thus, according to the model
derived in the main text, $p^>_Q(y)$ can be quantified by
\begin{align}
p^>_Q(y) \propto y^{2 \alpha/4 -1},
\end{align}
Figure \ref{fig_app:berea} shows the flow rate and velocity
distributions for the Berea pore network. We recover predicted scaling behaviors in
terms of the node degree and diameter distributions.

\section{Statistics}
\begin{figure*}[t!]
\includegraphics[width=0.4\textwidth]{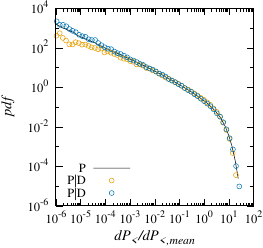}
\includegraphics[width=0.4\textwidth]{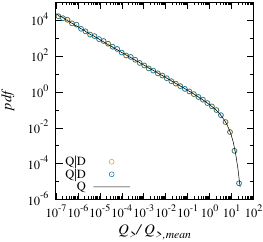}

\includegraphics[width=0.4\textwidth]{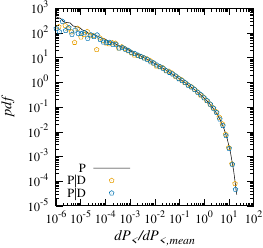}
\includegraphics[width=0.4\textwidth]{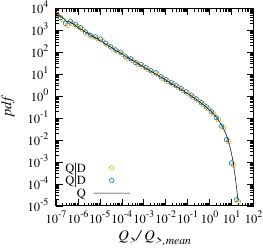}

\caption{Conditional (left) pressure statistics for $D < D_c$ for (orange) $D = 0.0004$ and
(blue) $D = 0.004$ and
(right) flow rate statistics for $D \geq D_c$ for (orange) $D = 0.08$ and
(blue) $D = 0.1$ for the (top row) honeycomb (bottom row) square
lattices with $\alpha = 1/2$. The conditional
statistics in the respective regimes collapse onto a single curve,
indicating their independence from diameter.
\label{fig_app:vsl}}
\end{figure*}
\subsection{Velocity distribution}

We consider the isochrone sampling of velocities along particle
trajectories. For generality, we assume that the conduit lengths are
variable and denoted by $\ell_{ij}$. The Lagrangian velocity
distribution is defined by isochrone sampling along the conduits
visited by a particle, and sampling over all particle
trajectories.

The routing of particles at a network node $i$ is according to the
downstream flow rate,  
\begin{align}
\label{eq:pij}
p_{ij} = \frac{|Q_{ij}|}{\sum\limits_{[ij)} |Q_{ij}|} = \frac{|Q_{ij}|}{\overline Q_i}.
\end{align}
The nodal flow rate is defined by $\overline Q_i = \sum_{[ij)}
|Q_{ij}|$, where $\sum_{[ij)}$ denotes
the sum over all nodes downstream from node $i$.
Particle velocities are sampled isochronically, that is, the nodal
probability is additionally weighted by the transition time
$\ell_{ij}/|v_{ij}|$ associated with the conduit. Thus, sampling
velocities isochronically at a node implies
\begin{align}
\nonumber
p_i(v) &\propto \sum_{[ij)} p_{ij} \frac{\ell_{ij}}{|v_{ij}|}
\delta(v - |v_{ij}|).
\end{align}
Finally sampling across all particle trajectories means weighting each
network node by its visitation frequency, 
which is proportional to $\overline Q_i$ such that $p_v(v) \propto
\sum_i \overline Q_i p_i(v)$. Thus, the Lagrangian
velocity distribution can be written as
\begin{align}
p_v(y) = \frac{\sum_i \overline Q_i \sum\limits_{[ij)}
\frac{|Q_{ij}|}{\overline Q_i}\frac{\ell_{ij}}{|v_{ij}|} \delta(y -
|v_{ij}|)}{\sum\limits_{i = 0}^{n} \overline Q_i \sum\limits_{[ij)}
\frac{|Q_{ij}|}{\overline Q_i}\frac{\ell_{ij}}{|v_{ij}|}}.
\end{align}
Using that $|Q_{ij}| = |v_{ij}| A_{ij}$, we obtain 
\begin{align}
p_v(y) = \frac{\sum_i \sum\limits_{[ij)}
A_{ij} \ell_{ij} \delta(y - |v_{ij}|)}{\sum_i \sum\limits_{[ij)} A_{ij} \ell_{ij}}. 
\end{align}
Setting $V_{ij} = A_{ij} \ell_{ij}$ and $V = \sum_i \sum\limits_{[ij)}
A_{ij} \ell_{ij}$, we obtain Eq.~\eqref{eq:pv} of the main text.  


\subsection{Conditional statistics}

\begin{figure}
\includegraphics[height=0.28\textwidth]{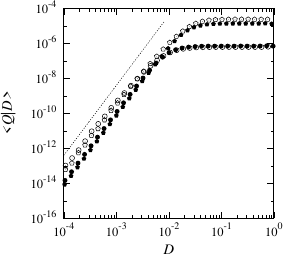}
\includegraphics[height=0.28\textwidth]{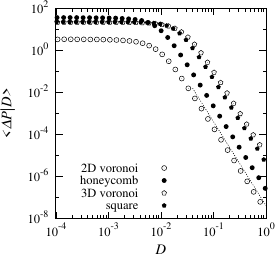}
\caption{Conditional averages of (left) flow rate and (right) pressure
drop for the (full symbols) regular and (open symbols) irregular
networks as functions of the conduit diameter for $\alpha =
1/2$. The dashed lines indicate the scalings $\langle Q|D\rangle
\propto D^4$, and $\langle \Delta P|D \rangle \propto
D^{- 4}$.\label{figapp:cond}}
\end{figure}

The distribution of flow rates conditional on the conduit diameter is
defined in terms of the joint PDF $p_{Q,D}(y,z)$ of flow rate and diameter as
\begin{align}
p_{Q|D_i}(y) = \int\limits_{D_i}^{D_i + \Delta D_i} d z p_{Q,D}(y,z)
\Big / \int\limits_{D_i}^{D_i + \Delta D_i} d z p_{D}(z)
\end{align}
and analogously for pressure drop and flow velocity. The conditional
mean flow rate is defined by
\begin{align}
\langle Q|D_i \rangle = \int\limits_0^\infty d y y p_{Q|D_i}(y). 
\end{align}
The conditional pressure and velocity statistics are defined
analogously. Figure \ref{fig_app:vsl} shows the conditional pressure
and flow rate distributions at $D < D_c$ and $D \geq D_c$,
respectively. For $D < D_c$ the conditional pressure distribution is
diameter-independent, whereas for $D \geq D_c$ the conditional flow rate
distribution is independent of the diameter. 
Figure \ref{figapp:cond} shows the conditional mean flow rate and mean
pressure drop for the networks under consideration. The conditional mean flow rate
behaves a $\langle Q|D\rangle \propto D^{4}$ for $D < D_c$ and
constant for $D \geq D_c$. The conditional mean pressure
drop behaves analogously and is independent of $D$ for $D < D_c$ and
decreases as $D^{-4}$ for $D \geq  D_c$.

\subsection{Critical diameter and percolation threshold}

The critical diameters $D_c$ are determined by fitting a spline to the
$\langle v|D \rangle$ data for the respective networks and
optimization of the resulting curve. The values for the proportion
$p_c$ of diameters $D \geq D_c$ are characteristic of the network types and
consistent with the bond percolation threshold. For the honeycomb lattice,
the percolation threshold is $0.652703645$, for the square lattice it
is $0.5$, for the simple cubic lattice, it is $0.2488126$, for the
2D Voronoi network it is $0.666931$ and for the 3D Voronoi network
$0.53618$ \cite{stauffer2018introduction,becker2009percolation}.   
 
\section{Pressure drop distributions}

The Dyson equation \eqref{eq:dyson} in the main text shows that the
pressure drop across a conduit depends in a non-local fashion on the
conductance fluctuations in the network, and integrates them in
general over an extended neighborhood. This observation explains why
the conditional pressure drop statistics in the $D < D_c$ cluster are
independent from the local diameter or conductance.

This can be illustrated by a first-order perturbation theory
calculation, which, albeit valid only for weak disorder, shows the
consequences of the non-local dependence of $\Delta P_{ij}$ on the
conductances. We define the conductance fluctuation by
\begin{align}
\kappa_{ij} = \kappa_0 + \kappa_{ij}',
\end{align}
where $\kappa_0 = \langle \kappa_{ij} \rangle$ is the mean conductance
and $\kappa_{ij}'$ the zero mean and finite variance
fluctuation. Furthermore, we
define the pressure fluctuations as $P_i'
= P_i - \Pi_i$, where $\Pi_i$ is the pressure solution for the
homogeneous network. The full Green function $G_{ij}$ satisfies
\begin{align}
\label{app:green}
\sum_{[ij]} \kappa_{ij} (G_{ik} - G_{jk}) = \delta_{ik},
\end{align}
where $\delta_{ij}$ is the Kronecker delta. The Dyson equation for the
Green function reads as
\begin{align}
G_{ij} = G^{(0)}_{ij} + \sum\limits_{k,\ell} G^{(0)}_{ik}
\kappa'_{k\ell} G_{\ell j},
\end{align}
where $G_{ij}^{(0)}$ solves \eqref{app:green} for $\kappa_{ij} =
\kappa_0$. 

\begin{figure*}[t!]

\includegraphics[width=0.4\textwidth]{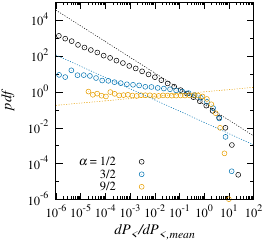}
\includegraphics[width=0.4\textwidth]{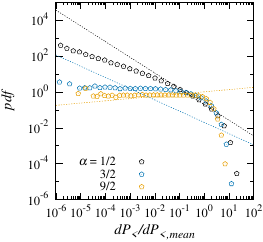}

\caption{Pressure drop distributions for (left) the honeycomb and
(right) the square lattice for different values of $\alpha$. The
pressure drop distributions for the Voronoi lattices behave in the same
way. The dashed lines show the power-law behaviors of the
corresponding conductance distributions $p_\kappa(k) \propto k^{\alpha/4-1}$. 
\label{fig:dp}}
\end{figure*}

For weak disorder, Equation \eqref{eq:dyson} for $\Delta P_{ij}$ can
be iterated by setting $G_{ij} = G^{(0)}_{ij}$, which gives
\begin{align}
\Delta P_{ij} = \Delta \Pi_{ij} + \sum_{k,\ell} \kappa_{k\ell}'
\left(G^{(0)}_{jk} - G^{(0)}_{ik}\right) \Delta \Pi_{k\ell}. 
\end{align}
The latter can be written as 
\begin{align}
\label{eq:dPij}
\Delta P_{ij} = \Delta \Pi_{ij} + \sum_{k,\ell} \kappa_{k\ell}' w_{ij,k\ell}, &&
w_{ij,k\ell} = \left(G^{(0)}_{ik} - G^{(0)}_{jk}\right) \Delta \Pi_{k\ell}. 
\end{align}
The weights $w_{ij,k\ell}$ fall off slowly with distance from the
conduit. Thus, the pressure drop is given by the weighted sum of
independent identically distributed random variables
$\kappa_{k\ell}'$, and the central limit theorem gives a
Gaussian distribution for the pressure drop
distribution. Specifically, this implies that $p_{\Delta P}(z)$ for $z
\ll \langle \Delta P \rangle$ behaves as $p_{\Delta P}(z) \propto
z^{0}$. This is not true anymore for strong disorder, that is, here
for conductance distributions with $\alpha/4 < 1$. Higher order terms
of the perturbation series become important. In fact, the pressure
drop distribution diverges for $z \to 0$ as a result of the high
frequency of low conductances. Nevertheless, and as a result of the
strong non-locality for strong disorder, the divergence of
$p_{\Delta P}(z)$ for $z \to 0$ is always weaker than
$z^{\alpha/4-1}$, which is the behavior of $p_\kappa(k)$, as a result
of the non-local dependence. While $p_{\Delta P}(z)$ does not show a
clear power-law behavior for $z \to 0$, we approximate it in the
following by $p_{\Delta P}(z) \propto z^{\delta-1}$, where $\alpha/4 <
\delta \leq 1$. The exponent is equal to $\delta = 1$ for weak disorder.  
Figure \ref{fig:dp} shows the distribution of the pressure drop
$|\Delta P|$ for the honeycomb and square networks for different
exponents $\alpha$. 

\section{Flow rate distributions}

In the following, we provide details for the calculation of the flow
rate distributions in the $D < D_c$ and $D \geq D_c$ clusters. 

\subsection{$D < D_c$}

The flow rate $|Q_{ij}|$ connecting nodes $i$ and $j$ is given by
\begin{align}
|Q_{ij}| = \kappa_{ij} |\Delta P_{ij}|, 
\end{align}
where $\kappa_{ij}$ is distributed according to
$p_\kappa(k)$. As the pressure drop is independent from $\kappa_{ij}$,
the flow rate distribution is given by
\begin{align}
\label{eq:pQ}
p_Q(y) = \int\limits_0^\infty dk \int\limits_0^\infty dz p_\kappa(k)
p_{\Delta P}(z) \delta(y - z k)
\end{align}
For $\alpha/4 \leq 1$, we perform the integration over $k$ to obtain
\begin{align}
p_Q(y) = \int\limits_0^\infty dz p_\kappa(y/z) z^{-1} p_{\Delta P}(z). 
\end{align}
We set $p_\kappa(k) \sim k^{\alpha/4-1}$ such that
\begin{align}
p_Q(y) \sim y^{\alpha/4-1} \int\limits_0^\infty dz z^{-\alpha/4} p_{\Delta P}(z). 
\end{align}
The integral on the right side converges because $\alpha/4 \leq 1$ and
the divergence of $p_{\Delta P}(z) \propto z^{\delta-1}$ with $\alpha/4 < \delta \leq 1$.
For $\alpha/4 > 1$ this is no longer true. In this case, we perform
the $z$-integration in \eqref{eq:pQ}, which gives
\begin{align}
p_Q(y) = \int\limits_0^\infty dk p_\kappa(k)
k^{-1} p_{\Delta P}(y/k). 
\end{align}
For small $y$, we approximate $p_{\Delta P}(z) \sim$ constant and thus
\begin{align}
p_Q(y) \sim \int\limits_0^\infty dk p_\kappa(k)
k^{-1} = \text{constant}. 
\end{align}
The integral exists because the integrand falls off exponentially fast
for large $k$ and behaves as $k^{\alpha/4 -2}$ for small $k$, where
$\alpha/4 - 2 > -1$. This explains (i) that the flow rate distributions scale as $p^<_Q(y)
\sim y^{\alpha/4-1}$ for $\alpha/4 \leq 1$ and as $p^<_Q(y) \sim$
constant for $y \ll \langle Q \rangle$, and (ii) that $p^<_Q(y)$ is
not sensitive to the network connectivity. 

\subsection{$D \geq D_c$}

In order to understand the behavior in the $D \geq  D_c$ clusters,
we distinguish between the well-connected flow paths in these
clusters and the network of conduits that connects them to dangling
ends which drain $(d-1)$ conduits that belong to the $D < D_c$
clusters. The flow rates $Q^>$ in these clusters scale with the
injection flow rate $Q_0^> = \sum_{i=1}^{d-1} Q^<_i$, which is
given by the sum of the flow rates $Q^<_i$ of the $(d-1)$ conduits with $D<
D_c$ connected to the dangling ends,
\begin{align}
Q^> = Q' Q_0^>. 
\end{align}
The dimensionless $Q'$ corresponds to the flow rate obtained for the
injection of a unit flow rate in the node connecting the dangling end
to the $D \geq D_c$ cluster. It integrates the impact of the conductance
fluctuations in an extended neighborhood, which for the $D \geq D_c$
clusters are governed by large conduits, or in other words, by a
narrow conductance distribution that is limited to the left by $D_c$
and to the right by the exponential cut-off. Thus, based on the perturbation
theory argument above for the pressure drop distribution, we assume
that the distribution $f_Q(y)$ of $Q'$ behaves as $f_Q(y) \propto y^0$ for $y \ll \langle Q
\rangle$ and falls off faster than exponential for large arguments. The
distribution of $Q^>$ is then given by
\begin{align}
\label{eq:pQlarge}
p_Q^>(y) = \int\limits_0^\infty dz p^>_{0}(z) z^{-1} f_Q(y/z).
\end{align}
The distribution $p^>_0(y)$ of $Q_0^>$ is obtained by $(d-1)$--fold convolution of
$p^<_Q(y)$. Thus, it scales as
\begin{align}
p_0^>(y) \propto y^{(d-1)\alpha/4 -1}, && p_0^>(y) \propto y^{d - 2}. 
\end{align}
for $\alpha/4 \leq 1$ and $\alpha/4 > 1$, respectively. Expression
\eqref{eq:pQlarge} guarantees that the flow rate distribution for $D >
D_c$ is indeed independent from the distribution of diameters, or
conductances.

In order to derive the scalings of $p_Q^>(y)$ for $y \ll \langle Q
\rangle$, we first consider the case $(d-1) \alpha/4 > 1$. In this
case, we set $f_Q(y/z) =$ constant and thus
\begin{align}
p_Q^>(y) \propto \int\limits_0^\infty dz p^>_0(z) z^{-1} =
\text{constant}. 
\end{align}
The integral on the right side converges because the integrand
behaves at small $z$ as $z^{(d-1)\alpha/4 -2}$, where $(d-1) \alpha/4
-2 > -1$ and at large $z$ it decreases faster than exponential. 
Second, we consider the case $(d-1) \alpha/4 \leq 1$. In this case,
we rewrite \eqref{eq:pQlarge} as
\begin{align}
p_Q^> = \int\limits_0^\infty dz' p^>_0(z' y) {z'}^{-1} f_Q(1/z'),
\end{align}
where we used the variable transform $z' = z/y$. For small $y \ll
\langle Q \rangle$, we set $p^>_0(z'y) \propto z'^{(d-1)\alpha/4-1}
y^{(d-1)\alpha/4-1}$ and so
\begin{align}
p_Q^> \propto y^{(d-1)\alpha/4-1} \int\limits_0^\infty dz'
z'^{(d-1)\alpha/4-2} f_Q(1/z') = \text{constant} \times y^{(d-1)\alpha/4-1}.   
\end{align}
The integral converges because the integrand behaves as
$z'^{(d-1)\alpha/4-2}$ with $(d-1)\alpha/4-2 < -1$ for $z' \gg 1$ and
it decays faster than exponential for $z' \to 0$. This model explains
(i) the independence of the flow rate statistics from the diameter in
the $D \geq  D_c$ clusters, (ii) the scaling of $p_Q^>(y) \propto
y^{(d-1)\alpha/4-1}$ for $(d-1)\alpha/4 < 1$ and (iii) $p^>_Q(y) =
\text{constant}$ for $(d-1) \alpha/4 > 1$ at $y \ll \langle Q
\rangle$. 

\end{document}